\documentclass[showpacs, prl, twocolumn, preprintnumbers, footinbib]{revtex4}

\usepackage{amsmath}
\usepackage{amssymb}
\usepackage{color}
\usepackage{verbatim}
\usepackage{graphicx}

\newcommand{\dif}{\mathrm{d}}

\newcommand{\mvac}{M_\text{vac}}
\newcommand{\tr}{\tilde{r}}
\newcommand{\tg}{\tilde{g}}

\newlength{\figwidth}
\begin{document}

\title{A monopole solution in a Lorentz-violating field theory}

\author{Michael D.\ Seifert}

\affiliation{Dept.\ of Physics, Indiana University, 727 E.\
$\text{3}^\text{rd}$ St., Bloomington, IN, 47405}

\email{mdseifer@indiana.edu}

\begin{abstract}
  I present a topological defect solution that arises in a theory
  where Lorentz symmetry is spontaneously broken by a rank-two
  antisymmetric tensor field, and I discuss its observational
  signatures.
\end{abstract}

\preprint{IUHET 546, August 2010}


\maketitle

Lorentz symmetry has been one of the cornerstones of physics for over
a century.  The possibility that this symmetry is approximate rather
than exact is an exciting prospect for the detection of physics beyond
the Standard Model, and has been the focus of many experimental
searches (see Ref.\ \cite{LVdata} and references therein.)  Such
symmetry breaking also arises in many theoretical models of current
interest, including string theory \cite{KosSam} and non-commutative
geometry \cite{LVNCgeom}.

A large class of theoretical models in which Lorentz symmetry is
broken involve a field which spontaneously breaks this symmetry
\cite{KosSam, Kosgrav, BKNG, LVAS}.  In such models, a vector or
tensor field is postulated to have a potential energy which is
minimized (and vanishes) at some field value other than zero.  This
behaviour is closely analogous to that of the Higgs field in the
Standard Model.  However, unlike the (scalar) Higgs field, a tensor
field with a non-zero expectation value will provide a preferred
geometric structure to Minkowski spacetime, and the vacuum solution
will not have the full symmetry of the Lorentz group. Direct couplings
between such a ``Lorentz-violating'' field and conventional matter
fields will then affect the dynamics of conventional matter, just as
direct Yukawa couplings between the Higgs field and fermion fields in
the Standard Model give mass to the fermion fields \cite{CKSME}.

An important class of solutions that arise in theories with a
spontaneously broken symmetry are topological defects.  These
solutions arise when the vacuum manifold of a symmetry-breaking field
(i.e., the set of field values which minimize the symmetry-breaking
potential) has certain non-trivial topological properties.  There can
then arise ``domain wall,'' ``cosmic string,'' or ``monopole''
solutions (respectively) to the equations of motion; on large scales,
these solutions effectively have two, one, and zero spatial
dimensions.  In theories in which topological defect solutions are
allowed, they will generically arise during phase transitions in the
early Universe, via the Kibble mechanism \cite{Kibble}.  Since
isolated defects are stable, it is possible that such relics still
exist in the Universe today.  The presence of topological defects can
also have important implications for cosmology, including inflation
\cite{GuthInflation} and structure formation \cite{TDstructure}.

The purpose of this letter is to show that topological defect
solutions can arise in theories with a tensor field that breaks
Lorentz symmetry.  Such solutions have not previously been described
in theories where the symmetry broken is Lorentz symmetry.  We will
derive an explicit example of such a solution, and discuss some of its
phenomenological properties.  

Our theory contains an antisymmetric two-tensor field (also known as a
notoph field \cite{Notoph} or Kalb-Ramond field \cite{KalbRamond})
which takes on a background expectation value \cite{LVAS}.  Its action
is of the form
\begin{equation}
  S_B = \int \dif^4 x \left( - \frac{1}{6} F^{abc} F_{abc} -
    \frac{\lambda}{2} (B^{ab} B_{ab} - b^2)^2 \right),
  \label{ASaction}
\end{equation}
where $B_{ab}$ is an antisymmetric tensor field and $F_{abc} =
3 \partial_{[a} B_{bc]}$ is its associated field strength \footnote{We
  will use a metric signature of $(-, +, +, +)$, and units where $c =
  h = 1$.}.  The equations of motion derived from this action are
\begin{equation}
  \label{ASEOM}
  \partial^c F_{cab} - 2 \lambda (B^{cd}
  B_{cd} - b^2) B_{ab} = 0 .
\end{equation}

Our first task is to show that the vacuum manifold of this theory has
the correct topology.  For a localized topological defect solution to
exist in three spatial dimensions, the vacuum manifold must either be
disconnected, contain a non-contractible loop, or contain a
non-contractible two-sphere; mathematically, it is necessary that one
of the vacuum manifold's homotopy groups $\pi_0(\mvac)$,
$\pi_1(\mvac)$, or $\pi_2(\mvac)$ be non-trivial.  From \eqref{ASEOM},
we can see that any constant field $B_{ab}$ such that $B_{ab} B^{ab} =
b^2$ is a solution of the equation of motion.  In terms of Cartesian
coordinate components, this condition is
\begin{equation}
  -2\sum_{i} (B_{0i})^2 + 2 \sum_{i < j} (B_{ij})^2 = b^2,
\end{equation}
where the summations run over spatial indices (i.e., $\{i, j\} =
{1,2,3}$.)  This condition defines a five-dimensional submanifold of
the six-dimensional field space of $B_{ab}$.  This submanifold is
homeomorphic to $S^2 \times \mathbb{R}^3$: for any choice of the three
components $B_{0i}$, the three components $B_{ij}$ are constrained to
lie on a sphere whose radius squared is $\frac{1}{2} b^2 + B_{0i} B_0
{}^i$.  Thus, the topology of the vacuum manifold allows for a
monopole solution, since it contains a non-contractible two-sphere
(i.e., $\pi_2(\mvac) = \mathbb{Z}$.)

These topological conditions on the vacuum manifold are necessary but
not sufficient for the existence of a monopole solution; we must still
find a solution of the equations of motion valid throughout space.  If
such solutions exist in this theory, we expect the simplest ones to be
static and spherically symmetric.  In standard spherical coordinates,
the most general antisymmetric two-tensor with this property can be
written in terms of two functions of $r$:
\begin{align}
  \label{Bcomps}
  B_{tr} = - B_{rt} &= f(r), & B_{\theta \phi} = - B_{\phi \theta} &=
  g(r) r^2 \sin \theta.
\end{align}
In terms of this ansatz, the equation of motion \eqref{ASEOM} has two
non-vanishing components, telling us that
\begin{subequations}
\begin{equation}
  \label{fvanish}
  -2 \lambda (- 2f^2 + 2g^2 - b^2) f = 0 
\end{equation}
and
\begin{equation}
  \label{gEOM}
  \frac{\partial}{\partial r}\left(
    \frac{\partial g}{\partial r} +
    \frac{2}{r} g \right) - 2 \lambda (-2 f^2 + 2 g^2 - b^2 ) g = 0.
\end{equation}
\end{subequations}
From the first of these equations \eqref{fvanish}, we see that either
$f$ must vanish or $B_{ab}$ must be in its vacuum manifold throughout
spacetime.  Since we only want our field to approach the vacuum
manifold asymptotically, we set $f = 0$.  Defining $\tr$ and
$\tg$ such that $g = b \tg / \sqrt{2}$ and $r = \tr/ (\sqrt{2
  \lambda} b)$, the second equation \eqref{gEOM} becomes
\begin{equation}
  \frac{\partial}{\partial \tr} \left(\frac{ \partial
      \tg}{\partial \tr} +
    \frac{2}{\tr} \tg \right) - (\tg^2 - 1) \tg = 0.
  \label{tgeqn}
\end{equation}

This equation is identical (up to rescaling) to that obeyed by
topological defects arising in a theory containing a triplet of
Lorentz scalars with a spontaneously broken global $O(3)$ symmetry
\cite{BarrVil}, for which a solution with $g(0) = 0$ and $g(r) \to
b/\sqrt{2}$ as $r \to \infty$ is known to exist.  No closed-form
expression for this solution is known, although series techniques
\cite{ShiLi} or numerical integration \cite{HarLou} can be used to
approximate it.  Asymptotically, the solution can be shown to obey
\begin{equation}
  \label{gasymp}
  \tg(\tr) = 1 - \frac{1}{2\tr^2} -\frac{3}{2 \tr^4} + \dots.
\end{equation}

In the absence of any direct coupling between $B_{ab}$ and
conventional matter fields, the principal method of detection of these
monopoles will be gravitational, via the coupling of their
stress-energy to the metric.  The stress-energy of the field $B_{ab}$
in a flat background is
\begin{multline}
  \label{ASSET}
  T_{ab} = F_{acd} F_b {}^{cd} - \frac{1}{6} \eta_{ab}
  F_{cde} F^{cde} \\ - \frac{\lambda}{2} \eta_{ab} (B^2 -
  b^2)^2  + 4 \lambda (B^2 - b^2) B_{ac} B_b
  {}^c,
\end{multline}
where $B^2 \equiv B_{ab} B^{ab}$.  

While the field profile of our tensor monopole is identical to that
of the $O(3)$ scalar monopole (as noted above), their respective
stress-energies differ in two important respects.  First, since the
kinetic term for $B_{ab}$ in the action \eqref{ASaction} is not simply
of a ``gradient-squared'' form, the kinetic terms in the stress-energy
tensor will differ from the analogous terms in the scalar monopole
stress-energy.  Second, the last term in Equation \eqref{ASSET} has no
analogue in the scalar case.  It arises from the differentiation of
the potential $V(B^2)$ with respect to the metric, since the
``square'' of a tensor field (unlike that of a scalar) depends on the
metric.

\begin{figure}
  \includegraphics[width=\columnwidth]{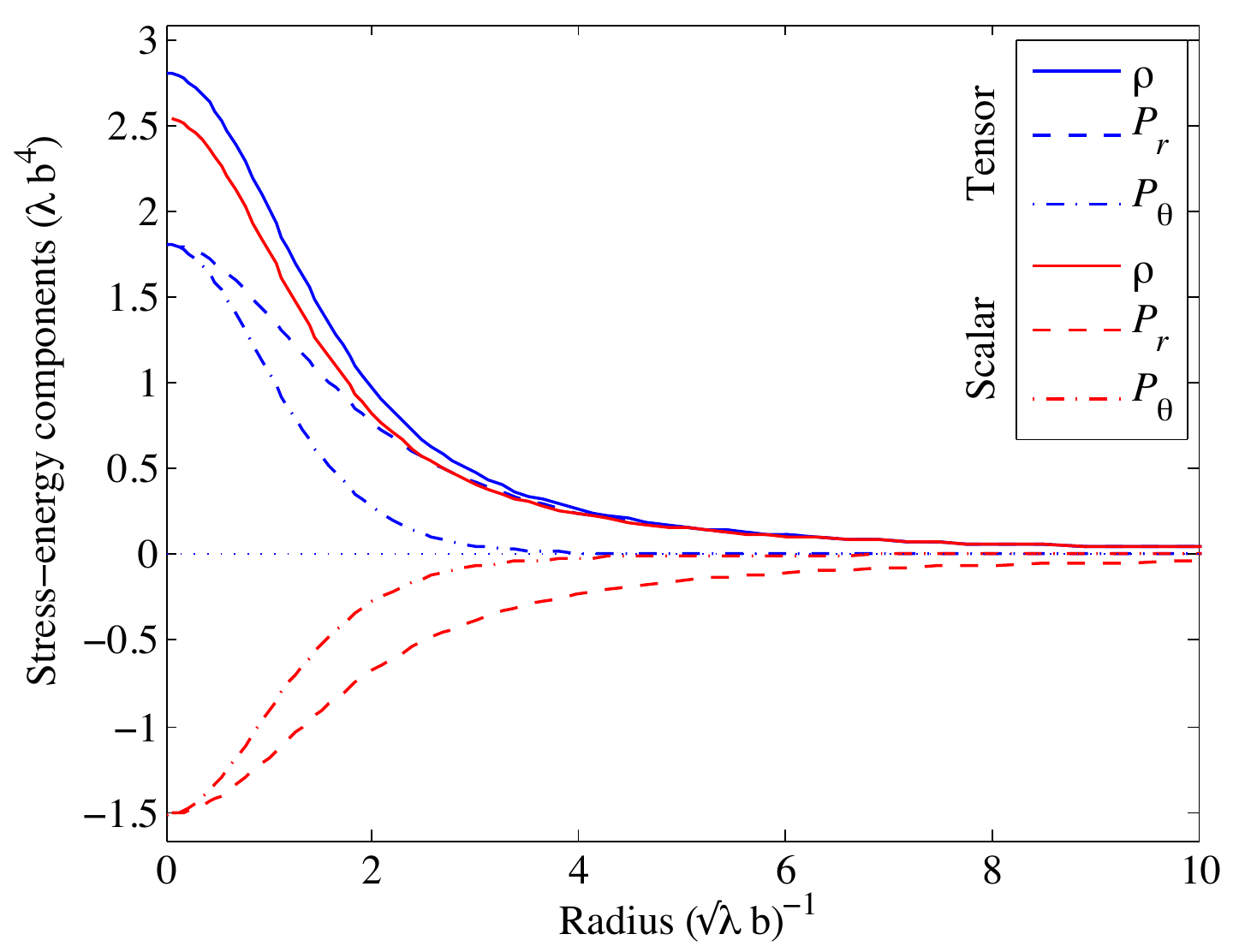}
  \caption{ \label{SEfig} Densities and pressures of the antisymmetric
    tensor monopole and the $O(3)$ scalar monopole.  Overall scaling
    for the $O(3)$ stress-energy is chosen so that the asymptotic
    fall-off of $\rho$ is the same in both solutions. }
\end{figure}
The components of the stress-energy for our tensor monopole are
compared to those of the $O(3)$ scalar monopole in Figure \ref{SEfig}.
Notably, while the radial and tangential pressures of the $O(3)$
monopole are negative, those of our tensor monopole are positive; this
difference is primarily due to the differences between the kinetic
terms of the two theories.  In flat spacetime, the leading-order
asymptotic behaviour of the energy density $\rho$, the radial pressure
$P_r$, and the tangential pressure $P_\theta$ can be shown to be
\begin{align}
  \rho \approx P_r &\approx \frac{4 \lambda b^4}{\tr^2}, & P_\theta
  \approx \frac{\lambda b^4}{\tr^4}.
\end{align}

To see the gravitational effects of our solution, we promote the
metric $g_{ab}$ to a dynamical field.  Assuming spherical symmetry and
staticity, we can write our line element as
\begin{equation}
  \label{metricform}
  ds^2 = - M^2(r) dt^2 + N^2(r) dr^2 + r^2 (d \theta^2 + \sin^2
  \theta d \phi^2 ).
\end{equation}
The independent components of the Einstein equation are then
equivalent to
\begin{subequations}
  \label{MNgEOMs}
\begin{multline}
  \frac{2}{r} \frac{N'}{N} + \frac{1}{r^2} (N^2 - 1) \\=
  \frac{\epsilon}{2 b^2} \left(
    \left( g' + \frac{2}{r} g \right)^2 +
    \frac{\lambda}{2} N^2 (2 g^2 - b^2)^2 \right) \label{EttEOM}
\end{multline}
and
\begin{multline}
  \frac{2}{r} \frac{M'}{M} - \frac{1}{r^2} (N^2 - 1) \\=
  \frac{\epsilon}{2 b^2}  \left( 
    \left( g' + \frac{2}{r} g \right)^2 - \frac{\lambda}{2} N^2 (2 g^2
    - b^2)^2 \right),  \label{ErrEOM}
\end{multline}
\end{subequations} 
where primes denote differentiation with respect to $r$ and $\epsilon
\equiv 16 \pi G b^2$. Meanwhile, the field equation of motion becomes
\begin{equation}
  \frac{N}{M} \frac{\partial}{\partial r} \left( \frac{M}{N} \left( g'
      + \frac{2}{r} g \right) \right) - 2 \lambda N^2 (2
  g^2 - b^2) g = 0
\end{equation}

While this system of equations is non-linear and rather intractable,
in the context of gravitational fields we are most interested in the
far-field effects. A useful approximation for investigating this regime in
monopole solutions (though one that must be used with some care in
this case \cite{Lortop}) is the BPS limit \cite{Bogo, PrasSomm}. In
this limit, we take $\lambda \to 0$ and look for solutions of the
Einstein equation \eqref{MNgEOMs} where the field takes on its
asymptotic value $g = b/\sqrt{2}$ everywhere. In this limit, the
equations \eqref{MNgEOMs} have the exact solution \begin{align}
  M^2(r) &= C_2 \frac{\tr^{1 + \epsilon} + C_1}{\tr^{1-\epsilon}} \label{Msol} \\
  N^2(r) &= (1 + \epsilon) \frac{ \tr^{1 + \epsilon}}{\tr^{1 +
      \epsilon} + C_1},
\end{align}
where $C_1$ and $C_2$ are constants of integration, the latter of
which can be set to unity via rescaling of $t$.

As noted above, the field profiles of our antisymmetric tensor
monopoles and those of global $O(3)$ monopoles are quite similar;
however, their gravitational fields have important qualitative
differences.  Both cases share the same asymptotic spatial geometry:
for both, the slices of constant $t$ are simply flat space with a
spherical deficit angle. However, unlike the scalar monopole case, the
component $g_{tt}$ in our solution grows without bound as $r \to
\infty$: from \eqref{Msol}, we see that $M(r) \propto
r^{\epsilon}$. This exceedingly slow divergence is due to the slow
fall-off of the stress-energy components, specifically the combination
$\rho + P_r + 2 P_\theta$ \footnote{This combination is proportional
  to the $tt$-component of the trace-reversed stress-energy tensor,
  which couples to the $tt$-component of the metric in the linearized
  approximation.}. Such a power-law divergence might indicate that the
solution for an isolated monopole is fundamentally non-static; this
would be analogous to anti-de Sitter space written in spherical
coordinates, where $g_{tt} \propto r^2$ asymptotically.  In a more
physically realistic situation, we only expect this solution to be
valid out to some finite radius, where the effects of larger structure
(on galactic or cosmic scales) take over. Since we expect to have
$\epsilon \ll 1$, the far-field geometry should be sufficiently flat
that we can ``patch'' our solution into one describing the appropriate
larger-scale structure.

The effects of this geometry on test particles, specifically test
photons, would be the primary method by which these monopoles could be
directly detected.  Two main effects on photons can be envisioned:
gravitational redshift and the bending of light rays.  The first of
these can be shown to be minimal: if our mass scale $b$ is well below
the Planck scale, the fractional gravitational redshift experienced by
a photon in this background will be within no more than two orders of
magnitude of $\epsilon$ \cite{Lortop}. Such effects will thus be
negligible.

A more interesting effect arises in the deflection of light rays in
this background. Using standard techniques, we can calculate that
to leading order a light ray propagating in this background will be
deflected by an angle \cite{Lortop} \begin{equation} \delta \phi
  \approx \frac{3 \pi}{2} \epsilon.
\end{equation}
Note that at this order, the deflection angle is independent of the
``apparent impact parameter.''  Rather, with respect to light bending,
the spacetime behaves as though it has a spherical deficit angle of
$\frac{3}{2} \epsilon$.  If a monopole were perfectly aligned between
an observer and a point source, the observer would see this point
source as a ring with angular diameter $\delta \phi \times l/(d+l)$,
where $d$ is the distance from the observer to the monopole and $l$
the distance from the monopole to the source.  A monopole slightly off
of the line of sight (but still sufficiently close to it) would give
rise to two images on the sky at this same angular separation.

From this point of view, the signature of an antisymmetric tensor
monopole is effectively the same as that of a global $O(3)$ monopole:
both give rise to apparent spherical deficit angles, albeit with
differing dependence on their respective mass scales.  Any
lensing-derived bounds placed on the current cosmological abundance of
global $O(3)$ monopoles will thus have a bearing on the abundance of
our tensor monopoles as well, and vice versa.

I am unaware of any direct observational searches specifically
targeted at this type of gravitational lensing.  However, a search for
angular deficits arising from cosmic strings, using images obtained as
part of the Great Observatories Origins Deep Survey, was recently
performed \cite{CSsearch}.  This search looked for morphologically
similar galaxies correlated on opposite sides of a line on the sky.
Similar techniques could be used to search for global monopoles; in
this case the doubled images would be correlated inside and outside of
a circle on the sky rather than on opposite sides of a line.  While
such structures are expected to be rare (see below), a search for
double images of distant galaxies would yield important information;
even if no monopoles were detected, one could in principle use a null
result from such a search to bound the mass scale $b$ associated with
our tensor field.

Other bounds on the abundance of $O(3)$ monopoles have also been
derived \cite{Hiscock, BennRhie}.  While such bounds are likely
adaptable to the present case, it is important to emphasize that they
are not immediately applicable in the same way that gravitational
lensing bounds would be.  The main reason for this difference is that
the stress-energy (and thus the far-field metric) of our tensor
monopole has important qualitative differences from that of the
$O(3)$ monopole; in particular, the combination $\rho + P_r + 2
P_\theta$ falls off as $r^{-2}$ for our tensor monopole, but as
$r^{-4}$ (due to cancellation of the $r^{-2}$ dependence) for the
$O(3)$ scalar monopole.

A final question is what the expected monopole density in the current
Universe should be.  While the Kibble mechanism predicts that one
monopole per Hubble volume (to within an order of magnitude) should
form during a phase transition in the early Universe, the subsequent
field dynamics might cause monopoles and antimonopoles to recombine in
the subsequent cosmological evolution. Many of the arguments and
counter-arguments concerning the recombination of global $O(3)$ scalar
monopoles \cite{BarrVil} would seem to apply here.  As in the scalar
monopole case, a definitive answer to this question is likely to
require computational simulation. 

Such simulations of scalar monopoles \cite{BennRhie} have found that
the density of such structures remains at approximately four monopoles
per Hubble volume throughout both the radiation- and matter-dominated
eras.  However, the above noted differences between the gravitational
fields of scalar monopoles and tensor monopoles might cause
recombination to proceed quite differently in the tensor case.  While
it still seems plausible that the results from scalar monopole
simulations would hold for our model as well (up to an order of
magnitude), such a statement should be regarded only as a conjecture.

If such a monopole were to be detected, it would obviously be of great
import to physics.  Such an observation would represent both the first
field observed (other than the metric) that is not described by the
Standard Model, and the first field observed to break Lorentz
symmetry.  As such, observation of a topological defect of this type
would provide invaluable insight into the roles played by fundamental
symmetries in physics.

Helpful discussions with B.\ Altschul, V.\ A.\ Kosteleck\'y, and M.\
Uhlmann concerning this work are gratefully acknowledged.  This
research was supported in part by the United States Department of
Energy under Grant No.\ DE-FG02-91ER40661.

\bibliographystyle{apsrev}
\bibliography{lortop}{}

\end{document}